\documentstyle[twocolumn,aps,prl,epsf]{revtex}
\newcommand{\bm}[1]{\mbox{\bf{#1}}}
\begin{document}
\draft
\title{{\em{Ab-initio}} calculation of the electronic and optical excitations in
polythiophene:\\
effects of intra- and interchain screening}
\author{J.-W.\ van der Horst, P.A.\ Bobbert, M.A.J.\ Michels}
\address{
Dept. of Applied Physics, COBRA Research School \& Dutch Polymer Institute,
Eindhoven University of Technology, \\P.O. Box 513, NL-5600 MB Eindhoven, The Netherlands}
\author{G.\ Brocks, P.J.\ Kelly}
\address{Dept. of Applied Physics, Twente University, P.O. Box 217, NL-7500 AE Enschede,
 The Netherlands
\begin{minipage}{5.5in} 
\begin{center}(\today)\end{center}
\begin{abstract}
We present an {\em ab-initio} calculation of
the electronic and optical excitations of an isolated polythiophene chain as well as of bulk
polythiophene. We use the $GW$ approximation for the electronic 
self-energy and include excitonic effects by solving the electron-hole
Bethe-Salpeter equation.
The inclusion of interchain screening in the case of bulk polythiophene drastically
reduces both the quasi-particle band gap and the exciton binding energies,
but the optical gap is hardly affected.
This finding is relevant for conjugated polymers in general.
\end{abstract}
\end{minipage} 
}
\maketitle

{\em{{Introduction --}}}
The interest in the electronic and optical properties of semiconducting conjugated polymers
has increased enormously since the discovery of
electroluminescence \cite{ppv} of these materials. 
The microscopic modelling of charge carriers and excitations in conjugated polymers is a subject 
of intensive research at present. In particular, knowledge of the single-particle bands and the 
electron-hole (exciton) binding energies is important. In conventional semiconductors such as GaAs
the optical excitations are well described in terms of very weakly bound electron-hole pairs (Wannier 
excitons). In crystals made of small organic molecules such as anthracene, the electron-hole correlation 
energy is large, which essentially confines the exciton to a single molecule (Frenkel exciton).
Exactly where conjugated polymers fit in between conventional semiconductors on the one hand and molecular
crystals on the other, is the subject of a heated debate \cite{bredasheeger}. Most of the experimental 
results have been
obtained from optical experiments on the condensed (thin film or bulk) material.
Most of the theoretical modelling has been at the level of a single molecule or a single polymer chain.
In this paper we will focus on the differences between an isolated polymer chain and a bulk polymer 
material and show that these have a large effect on basic electronic parameters such as the band gap and
the exciton binding energies. We calculate the electronic structure
and the optical excitation of conjugated polymers in an {\em{ab-initio}} manner, 
i.e.\ without `tunable' parameters. 

{\em{Ab-initio}} calculations of several polymers within the Local Density Approximation of
Density Functional Theory (DFT-LDA) yield band gaps that are typically 40\% smaller than the 
experimental optical gaps~\cite{bkc}. However, it is well known that DFT-LDA formally cannot 
be applied to excited states.
Moreover, no excitonic effects are taken
into account in these calculations.
An {\em{ab-initio}} many-body $GW$ approximation ($GW$A)~\cite{gw}
calculation on polyacetylene (PA) was performed by Ethridge {\em{et al.}}~\cite{ethridge},
but excitonic effects were not taken into account 
in their work either. Very recently, a calculation by Rohlfing and Louie \cite{oeicp}
for {\em{isolated}}
chains of both PA and poly-phenylene-vinylene (PPV),
including both single- (quasi-particle) and two-particle (exciton) excitations,
yielded optical gaps in good agreement with experiment. However, 
the calculated lowest-lying singlet exciton binding energy for PPV in that
work is much larger than in a recent experiment~\cite{intermediate}.

In this article, we calculate single- and two-particle excitations for
an isolated polythiophene (PT) chain {\em{as well as}} for bulk PT, where we will show that
the differences between the two cases have very important consequences.
We focus on PT, since it is one of the simplest and most studied polymers.
Our main conclusions, however, are relevant for conjugated polymers in general.

{\em{{Computational Methods --}}} 
Many successful ab-initio calculations of the quasi-particle (QP)
band structure of conventional
anorganic semiconductors have been performed with the $GW$A
for the electronic self-energy $\Sigma$ of the one-particle Green function.
Very recently~\cite{ar,rl}, important progress has been made in the evaluation of the two-particle Green function, 
from which the optical properties can be obtained. This is done by solving the
Bethe-Salpeter equation\cite{shamrice,strinati} (BSE), which is equivalent to a two-body 
Schr\"odinger equation, for an electron and a hole forming an exciton. We adopt these two
methods and neglect the differences in geometry between the excited states and the
ground state. DFT-LDA calculations on oligothiophenes show that relaxation energies,
calculated by relaxing an excited state from the ground state geometry to its optimized
geometry, typically are in the range 0.1-0.2 eV \cite{oligothiophene}. Such values are
upperbounds for the corresponding states of the polymer. 

We start our calculations with a pseudo-potential plane-wave DFT-LDA
calculation~\cite{bkc} of a geometry-relaxed 
PT chain in a tetragonal supercell (14.8 a.u.\ in the chain direction, 15 a.u.\ in the
perpendicular directions, the latter value large enough to consider the chains as isolated in
this calculation).
We use Hartree atomic units unless specified otherwise.
The single-particle excitation energies $E_{nk}$ are evaluated by solving the QP equation
\begin{eqnarray}
&&\left[-\frac{1}{2}\nabla^2 + V_{H}(\bm{r}) \right]\phi_{nk}(\bm{r}) +\nonumber\\
&&\int \left[V_{PP}(\bm{r}, \bm{r}') + \Sigma(\bm{r}, \bm{r}', E_{nk})\right]
\phi_{nk}(\bm{r}')d^3r'
 = E_{nk} \phi_{nk}(\bm{r}),
\end{eqnarray}
where $V_{PP}$ is the pseudo-potential of the atomic core, $V_H$ the Hartree potential due to the 
valence electrons,
and $\Sigma$ the self-energy. We make the usual approximation that the QP wave functions
$\phi_{nk}$ can be replaced by the DFT-LDA wave functions.
In DFT-LDA, $\Sigma(\bm{r}, \bm{r}', \omega)~=~V_{xc}(\bm{r})\delta(\bm{r}-\bm{r}')$, with
$V_{xc}(\bm{r})$ the exchange-correlation potential,
while in the $GW$A
the first term in the many-body expansion of $\Sigma$
in terms of the Green function $G$ and screened interaction $W$ of the system is used,
$\Sigma(\bm{r}, \bm{r}',t)=i G(\bm{r}, \bm{r}',t)W(\bm{r}, \bm{r}',t)$, after Fourier
transformation from frequency to time.
The screened interaction is evaluated in the random-phase approximation (RPA).
We calculate these approximations of $\Sigma$ and $W$ in the mixed-space
imaginary-time formalism~\cite{spacetime,msform}, from the DFT-LDA wave functions and energies.
To study a truly isolated chain, we remove 
the `crosstalk' between periodic images of the chain by
taking as a unit cell the spatial region
closer to the atomic positions of a specific chain than to those of any other;
the Coulomb interaction $v(\bm{r}-\bm{r}')=1/|\bm{r}-\bm{r}'|$ is then cut off
by setting it zero if 
$\bm{r}$ and $\bm{r}'$ belong to different cells. 

The two-body electron-hole Schr\"odinger equation related to the BSE is solved by expanding the 
exciton wave functions $\Phi(\bm{r}_e, \bm{r}_h)$ of zero-momentum excitons, which are the
optically relevant ones,
in products of conduction ($c$)
and valence ($v$) wave functions,
$\Phi(\bm{r}_e, \bm{r}_h) = \sum_{kcv} A_{kcv}\phi_{ck}(\bm{r}_e)\phi^*_{vk}(\bm{r}_h)$.
The expansion coefficients $A_{kcv}$ and the exciton binding energy $E_b$ 
should obey~\cite{ar,rl,shamrice,strinati}:
\begin{eqnarray}
&&[E_{ck} - E_{vk} - E_g + E_b ] A_{kcv} 
+ \nonumber\\
&&\sum_{k'c'v'} [2V^x_{kcv,k'c'v'}-W_{kcv,k'c'v'}]A_{{k'c'v'}} = 0,
\end{eqnarray}
where $E_g$ is the QP band gap, $E_{ck}$ and $E_{vk}$ are the QP energies, 
and $W_{kcv,k'c'v'}$ and
$V^x_{kcv,k'c'v'}$ (only present for singlet excitons)  
are the matrix elements of the static ($\omega = 0$)
screened interaction and 
the exchange matrix elements of the bare Coulomb interaction, respectively. 
Formally, dynamical
screening effects may only be ignored in the BSE if $E_g \gg E_{b}$.
However, it has been shown that
dynamical effects in the electron-hole screening and in the
one-particle Green function largely cancel each other \cite{bechstedt} and hence
the static approximation
is expected to be accurate, even when the relation $E_g \gg E_{b}$ does not strictly hold.

As was shown by several authors~\cite{herman,schulz2}, 
in a quasi-1D system (such as an isolated polymer chain with only
intrachain screening)
there is no long-range screening. For realistic bulk polymers, however, both
intra- and interchain screening are important, the latter giving
rise to the screening at long distances.
While it is in principle possible to perform a $GW$A and exciton calculation for a crystalline
structure, the amount of computational work is as yet prohibitively large.
Instead, we choose to include the interchain screening in the following way, capturing the
essential physics.
From the imaginary-frequency dependent
RPA polarizability tensor of an isolated chain in the long-wavelength limit
we obtain a dielectric constant parallel ($x$) to the chain direction, $\varepsilon_{||}(i\omega)$,
and perpendicular ($y$,$z$) to the chain direction, $\varepsilon_{\bot}(i\omega)$,
by performing a 2D line-dipole lattice sum in the perpendicular directions, 
for an experimentally determined PT crystal structure~\cite{mo,xtal}. 
Details of this approach
will be given in a future article~\cite{onszelf}.
We now take for the interchain screening interaction $W^{\rm scr}\equiv W-v$ the
following form:
\begin{eqnarray}
&&W^{\rm scr}_{\rm{inter}}(\bm{r},i\omega) = (1-e^{-r/r_{\rm{inter}}})\times\nonumber\\  
&&\left\{ \left[
         \varepsilon_\bot^2(i\omega)x^2+
\varepsilon_{||}(i\omega)\varepsilon_\bot(i\omega)(y^2+z^2)\right]^{-1/2} -v(\bm{r}) \right\},
\end{eqnarray}
which is correct for distances much larger than the typical
interchain distance $r_{\rm{inter}}$, for which we take 10~a.u.\
(typical for the crystal structure of Refs.~\onlinecite{mo,xtal}). 
The prefactor in this equation takes care of a soft cutoff for distances below
$r_{\rm{inter}}$,
for which the interchain screening should vanish.
We remark that the details of this interaction cannot
be important since PT samples prepared in many different ways show very similar
optical behavior; probably, most samples consist of small crystallites separated by
disordered regions.
We now construct a total screened interaction
\begin{eqnarray}
W_{\rm{total}}(\bm{r},\bm{r}',i\omega) &=& W^{\rm scr}_{\rm{intra}}(\bm{r},\bm{r}',i\omega)+
\nonumber\\&&
W^{\rm scr}_{\rm{inter}}(\bm{r}-\bm{r}',i\omega) +v(\bm{r}-\bm{r}'),\label{total}
\end{eqnarray}
where $W^{\rm scr}_{\rm{intra}}$ is the screening interaction already calculated for
the isolated chain.
The interaction Eq.~(\ref{total})
is correct at short and long range and is expected to be sufficiently good at
intermediate range.
We remark that
the overlap between wave functions on, and therefore the {\em{electronic}} coupling between,
neighboring chains is very small; hence, we can again use the isolated chain wave functions 
to calculate
the Green function and self-energy. In our exciton calculations, we take the electron and hole 
on the same chain (so-called intrachain excitons). Hence, the only,
but essential, difference
between our calculations for the isolated PT chain and bulk PT is in the screened
interaction used.

We determine the various cutoff parameters in our calculations (number of $k$-points,
bands, grid mesh in space and 
imaginary time) such that the QP band gap is converged to within 0.05 eV.
In the construction of the exciton wave functions 
we use only the $\pi$ and $\pi^{*}$ wave functions, resulting in an accuracy of 
the lowest-lying exciton binding energies of about 0.1 eV.
Below we give all calculated energies in eV with a precision of two decimal places.

\begin{figure}
\leavevmode
\centerline{
\epsfxsize\linewidth
\epsffile[60 50 350 500]{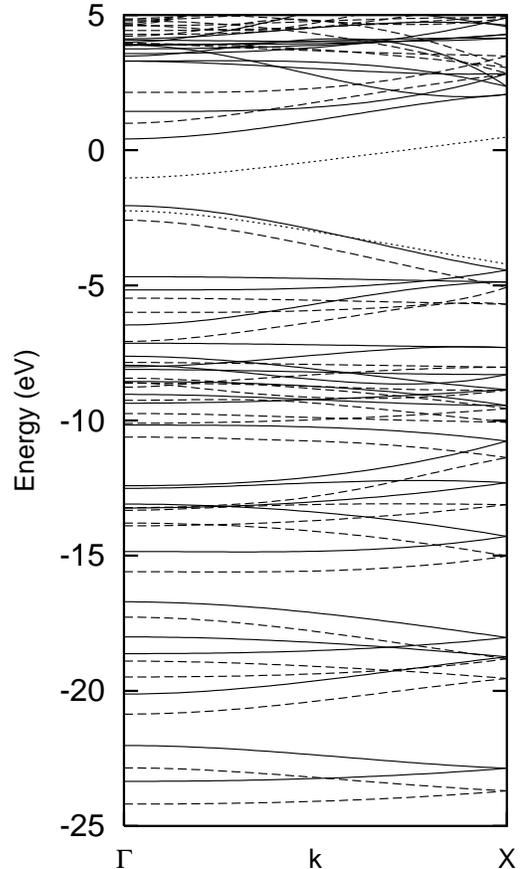}}
\caption{The single-particle band structure of the valence electrons of polythiophene along
the chain direction
calculated within the $GW$A, including intrachain screening
only (isolated chain, dashed) and including both intra- and interchain screening
(bulk, full lines). The 
DFT-LDA highest valence ($\pi$) and lowest conduction ($\pi^{*}$) bands are drawn dotted.}
\label{qp}
\end{figure}

{\em{{Results --}}}
The calculated $GW$A QP band structure of the isolated chain is shown in Fig.~\ref{qp}. 
We find a
band gap of 3.59 eV (DFT-LDA: 1.22 eV). 
The binding energies of the lowest-lying excitons and the resulting optical band gap
are listed in Table \ref{excagain}.  
The lowest-lying singlet exciton ($^1{\rm{B}}_{\rm{u}}$) for the isolated chain
has a binding energy of 1.85 eV,
leading to an optical gap of 1.74 eV. To our knowledge, no direct 
experimental information is available for
either the QP band gap or the $^1{\rm{B}}_{\rm{u}}$ exciton binding energy of PT separately.
Only the difference,
i.e.\ the optical 
gap, corresponding to the onset of the optical absorption in a well-ordered alkyl-substituted
polythiophene, is known to be about 1.8 eV \cite{mccullough,sakurai}. This is in good agreement with our calculated
optical gap. Similar conclusions regarding the optical gap were reached by
Rohlfing and Louie \cite{oeicp} for isolated PA and PPV chains.

However,
the {\em difference} between the $^1{\rm{B}}_{\rm{u}}$ and $^1{\rm{A}}_{\rm{g}}$
binding energies of the isolated PT chain is definitely
{\em{not}} in agreement with a recent experiment~\cite{sakurai}, see Table~\ref{excagain}. 
Moreover, the $^1{\rm{B}}_{\rm{u}}$ exciton binding energy of 1.85 eV is very
large compared to values currently discussed in the literature.
There is a heated debate going on
about the magnitudes of exciton binding energies in conjugated polymers \cite{bredasheeger}.
Negligibly small (0.1 eV or less~\cite{onetenthorless}), intermediate 
($\sim$ 0.5 eV~\cite{intermediate}), 
and large ($\sim$ 1.0 eV~\cite{oneev}) energies have been proposed.
However, these values concern either films or bulk systems,
both of which are essentially 3D. 

The QP band structure calculated with our model 3D interaction Eq.~(\ref{total})
for bulk PT
is also given in Fig.~\ref{qp}. The band gap has
decreased to 2.49 eV. The results for the exciton binding energies are included in 
Table~\ref{excagain}.
The $^1{\rm{B}}_{\rm{u}}$ exciton binding energy has decreased to 0.76 eV, resulting in an
optical gap of 1.73 eV. So, even though both the exciton binding energy and the
band gap have changed considerably, the optical gap has hardly changed compared to the 
isolated chain. 
Furthermore, the relative exciton energies are now in good
agreement with the experimental data~\cite{sakurai}. 
If we fit our exciton coefficients to 
the hydrogen-like form $A_k/A_{k=0} = 1/(1+a^2_{\rm{ex}}k^2)^2$, we obtain for the
$^1{\rm{B}}_{\rm{u}}$ exciton sizes $a_{\rm{ex}}\sim 12$~a.u.\ and $\sim 18$~a.u.\
in the case of the isolated chain and the bulk situation, respectively. 

In order to test the sensitivity of our results to the precise value of the cutoff
distance $r_{\rm{inter}}$ we performed similar calculations for $r_{\rm{inter}}$~=~8~a.u.\
and $r_{\rm{inter}}$~=~12~a.u. The QP band gaps are 2.32 and 2.61~eV, respectively. 
The $^1{\rm{B}}_{\rm{u}}$ binding energies are 0.64 and 0.86~eV and hence the
optical gaps are 1.68 and 1.73~eV, respectively. This means that the optical gap
is quite insensitive to the choice of $r_{\rm{inter}}$. The energy differences between
the excitons are also hardly influenced.

{\em{Discussion and conclusions --}}
Summing up, we have calculated the 
single-particle band structure and lowest-lying exciton binding energies of an isolated
polythiophene chain and bulk polythiophene.
For the isolated chain (only intrachain screening) we find a large band gap and
large exciton binding energies, due to the absence of long-range screening. 
Upon including interchain screening, responsible for the long-range screening
present in bulk polythiophene,
we find that both the band gap and exciton binding energies 
are drastically reduced. However, the optical gap hardly changes. 

We suggest that these conclusions hold for conjugated polymers in general.
In this light, we can understand the fact that the calculations by Rohlfing and Louie~\cite{oeicp}
on isolated chains of polyacetylene and PPV yield good results for the optical gaps, whereas their
lowest-lying singlet exciton binding energy of 0.9 eV for PPV is far in excess of a recent 
experimental value of $0.35\pm 0.15$~eV~\cite{intermediate}, obtained by a direct
STM measurement. The inclusion of interchain screening effects will drastically reduce
the calculated binding
energy (by more than a factor of two in our case of polythiophene), and may well lead to better
agreement with this experiment.
A similar drastic reduction of the exciton binding energy  by interchain screening
effects was predicted recently by Moore and Yaron~\cite{mooreandyaron} in polyacetylene,
within the semi-empirical Pariser-Parr-Pople theory.  
Clearly, it would be very
interesting
to repeat the experiment in Ref.~\onlinecite{intermediate} for polythiophene and polyacetylene.

We conclude that a correct many-body description of the electronic properties of
bulk polymer systems should include the effect of interchain screening. 
An important consequence is that neither Hartree-Fock nor DFT-LDA calculations 
should be relied upon in this context, since Hartree-Fock does not contain
screening effects at all and since the exchange-correlation
potential in DFT-LDA only depends on the local density and cannot 
describe the non-local effects due to the long-range screening.

Financial support from NCF (Dutch National Computing Facilities) through project SC-496 is
acknowledged. G. Brocks acknowledges the financial support from Philips Research through
the FOM-LZM (fundamenteel onderzoek der materie - laboratorium zonder muren) program.
We thank Michael Rohlfing for very fruitful discussions.

\begin{table}
\begin{tabular}{l r@{.}l r@{.}l r@{/}l}
 & \multicolumn{2}{c}{intra} &  \multicolumn{2}{c}{intra$+$inter}
 & \multicolumn{2}{c}{experiment} \\
\hline
$^1{\rm{A}}_{\rm{g}}$ & 0&96 &  0&23 &  \multicolumn{2}{c}{ }  \\
$^3{\rm{A}}_{\rm{g}}$ & 1&11 &  0&32  &\multicolumn{2}{c}{ } \\
$^1{\rm{B}}_{\rm{u}}$ & 1&85 &  0&76 &  \multicolumn{2}{c}{ }\\
$^3{\rm{B}}_{\rm{u}}$ & 2&36  & 1&15 &  \multicolumn{2}{c}{ } \\
\multicolumn{7}{c}{ } \\
$^3{\rm{B}}_{\rm{u}} \rightarrow\ ^1{\rm{B}}_{\rm{u}}$
& 0&51 & 0&39 & 0.45&0.45 \\
$^1{\rm{B}}_{\rm{u}}  \rightarrow\ ^1{\rm{A}}_{\rm{g}}$
& 0&89 & 0&53 & 0.50&0.55 \\
\multicolumn{7}{c}{ } \\
$E_g$
& 3&59 & 2&49 & \multicolumn{2}{c}{ }  \\
$E_o$
& 1&74 & 1&73 & 2.0&1.8

\end{tabular}
\caption{Binding and transition 
energies of the four lowest-lying excitons in polythiophene, and quasi-particle 
($E_g$) 
and optical ($E_o$) band
gaps
calculated with
intrachain screening only (isolated chain) and intra- plus interchain
screening (bulk). Experimental data before/after lattice relaxation according to
Ref.~\protect\onlinecite{sakurai}. All in eV.}
\label{excagain}
\end{table}


\begin{thebibliography}{99}

\bibitem{ppv}
J.H.\ Burroughes, D.D.C.\ Bradley, A.R.\ Brown, R.N.\ Marks, K. Mackay,
R.H.\ Friend, P.L.\ Bum, A.B.\ Holmes, Nature {\bf{347}}, 359 (1990).

\bibitem{bredasheeger}
J.-L. Br\'edas, J. Cornil, A.J. Heeger, Adv. Mat. {\bf{8}}, 447 (1996).

\bibitem{bkc}
G. Brocks, P.J. Kelly, R. Car, Synth.\ Met.\ {\bf{55-57}}, 4243 (1993).

\bibitem{gw} L. Hedin, Phys.\ Rev.\ {\bf 139}, A796 (1965); for a review of the $GW$A see:
F. Aryasetiawan and O. Gunnarsson, Rep.\ Prog.\ Phys.\ {\bf 61}, 237 (1998).

\bibitem{ethridge}
E.C.\ Ethridge, J.L.\ Fry, M.\ Zaider, Phys.\ Rev.\ B {\bf{53}}, 3662 (1996).

\bibitem{oeicp}
M. Rohlfing and S.G. Louie, Phys.\ Rev.\ Lett.\ {\bf{82}}, 1959 (1999).

\bibitem{intermediate}
S.F.\ Alvarado, P.F.\ Seidler, D.G.\ Lidzey, D.D.C.\ Bradley, Phys.\ Rev.\ Lett.\ {\bf{81}}, 1082 (1998).

\bibitem{ar} 
S. Albrecht, L. Reining, R. Del Sole and G. Onida,
Phys.\ Rev.\ Lett.\ {\bf{80}}, 4510 (1998).

\bibitem{rl}
M. Rohlfing and S.G. Louie, Phys.\ Rev.\ Lett.\ {\bf{81}}, 2312 (1998).

\bibitem{shamrice}
L. Sham and T.M. Rice, Phys.\ Rev.\ {\bf{144}}, 708 (1965).

\bibitem{strinati}
G.\ Strinati, Phys.\ Rev.\ B {\bf{29}}, 5718 (1984).

\bibitem{oligothiophene} 
The properties of excited states in finite systems such as oligothiophenes can be 
calculated within the DFT-LDA formalism, provided they have a symmetry different from that of
the ground state. For instance, the relaxation energy associated with the $^3{\rm{B}}_{\rm{u}}$ state of 
12T (an oligothiophene consisting of 12 thiophene rings) is 0.2 eV. Singlet states typically
have a smaller relaxation energy.

\bibitem{spacetime}
H.N. Rojas, R.W. Godby, and R.J. Needs, Phys. Rev. Lett. {\bf{74}}, 1827 (1995).

\bibitem{msform}
X.\ Blase, A. Rubio, S.G.\ Louie, M.L.\ Cohen, Phys.\ Rev.\ B {\bf{52}}, R2225 (1995).

\bibitem{bechstedt}
F.\ Bechstedt, K.\ Tenelsen, B.\ Adolph, R.\ Del Sole,
Phys.\ Rev.\ Lett.\ {\bf{78}}, 1528 (1997).
 
\bibitem{herman}
H.J. de Groot, P.A. Bobbert, and W. van Haeringen, Phys.\ Rev.\ B {\bf{52}}, 11000 (1995).

\bibitem{schulz2}
H.J.\ Schulz, Phys.\ Rev.\ Lett.\ {\bf{71}}, 1864 (1993) and references therein.

\bibitem{mo}
Z.\ Mo, K.-B.\ Lee, Y.B.\ Moon, M.\ Kobayashi, A.J.\ Heeger, F.\ Wudl, Macromol. {\bf{18}}, 1972 (1985).

\bibitem{xtal}
S. Br\"uckner and W. Porzio, Makromol. Chem. {\bf{189}}, 961 (1988).

\bibitem{onszelf}
J.-W. van der Horst, P.A.\ Bobbert, M.A.J.\ Michels, G.\ Brocks, P.J.\ Kelly,
in preparation.

\bibitem{mccullough}
R. D. McCullough, R. D. Lowe, M. Jayaraman, and D. L. Anderson,
J. Org. Chem. {\bf 58}, 904 (1993).

\bibitem{sakurai}
K. Sakurai, H. Tachibana, N. Shiga, C. Terakura, M. Matsumoto and Y. Tokura,
Phys. Rev. B {\bf{56}}, 9552 (1997).

\bibitem{onetenthorless}
T.W.\ Hagler, K.\ Pakbaz, A.J.\ Heeger, Phys.\ Rev.\ B {\bf{49}}, 10968 (1994).

\bibitem{oneev}
M.\ Chandross, S.\ Mazumdar, S.\ Jeglinski, X.\ Wei, Z.V.\ Vardeny, E.W.\ Kwock, T.M.\ Miller,
Phys.\ Rev. B {\bf{50}}, 14702 (1994).

\bibitem{mooreandyaron}
E.E.\ Moore and D.\ Yaron, J. Chem. Phys. {\bf{109}}, 6147 (1998).

\end{thebibliography}
\end{document}